\documentclass[aps,prb,onecolumn,11pt,tightenlines,superscriptaddress,
               amsmath,amssymb]{revtex4}
\usepackage{graphicx}
\usepackage{dcolumn}

\begin{document}

\title{Two exchange-correlation functionals compared for
first-principles liquid water}

\author{M. V. Fern\'andez-Serra}
\email{mfer01@esc.cam.ac.uk}
\affiliation{Department of Earth Sciences,
             University of Cambridge,
             Downing street, Cambridge CB2 3EQ, UK}
\author{G. Ferlat}
\affiliation{Laboratoire de Min\'eralogie-Cristallographie,
             Universit\'e Pierre et Marie Curie, Paris, France}
\author{Emilio Artacho}
\affiliation{Department of Earth Sciences,
             University of Cambridge,
             Downing street, Cambridge CB2 3EQ, UK}

\date{27 July 2004}

\begin{abstract}
  The first-principles description of liquid water using  
ab initio molecular dynamics (AIMD) based on Density Functional
theory (DFT) has recently been found to require long equilibration 
times, giving too low diffusivities and a clear over-structuring of 
the liquid.
  In the light of these findings we compare here the room-temperature
description offered by two different exchange correlation functionals:
BLYP, the most popular for liquid water so far, and RPBE, a 
revision of the widely used PBE.
  We find for RPBE a less structured liquid with radial
distribution functions closer to the experimental ones than the
ones of BLYP. 
  The diffusivity obtained with RPBE for heavy water is still
20\% lower than the corresponding experimental value, 
but it represents a substantial improvement on the BLYP value,
one order of magnitude lower than experiment.
  These characteristics and the hydrogen-bond (HB) network imperfection
point to an effective temperature $\sim$3\% lower than the
actual simulation temperature for the RPBE liquid, as compared 
with BLYP's $\sim$17\% deviation.
  The too long O--O average nearest-neighbor distance observed 
points to an excessively weak HB, possibly compensating more
fundamental errors in the DFT description.
\end{abstract}

\maketitle

\section{Introduction}

  The interaction of organic pollutant molecules with the minerals and 
other components of the soil represents a paradigmatic case within
the set of atomic-scale problems of relevance for the environment. 
  It displays the level of complexity that demands a fundamental 
revision of the way we use computers for the simulation of condensed matter.
  On one hand the number of adsorption studies to address is enormous
(combinations of many different pollutants, minerals, surfaces thereof etc.),
on the other, each study displays a large degree of complexity, not least 
because of the wet character of these systems.

  The effect of water in this and other wet systems is crucial
in different ways, from the dielectric screening of
electrostatic interactions to the active chemical role played
in some important wet processes.
  These different aspects of water can be described at
different levels of theory, including the static calculation
of the system of interest immersed in an effective dielectric
continuum, or the dynamic simulation of the liquid based on
empirical potentials of different kinds.
  Even if simplified descriptions of water can be appropriate
for some problems, other problems require the accuracy and 
transferability provided by first-principles simulation methods.
  This is not only our case, but it is true for many wet systems
throughout geochemistry, biochemistry and wet chemistry. 
  These considerations and the high variety of situations that we
encounter in our pollutant problem (see Ref.~\onlinecite{clovis} for the 
combinatorial aspects of the project and the workflow issues it
entails) make it extremely important that we can use different simulation
methods that involve different computational demands in a seamless 
fashion and within a single computational 
environment.\cite{condor,AllHands,JonXML,grid}
  We need not only to interconnect the different simulations\cite{Jon} 
but that each one of them performs at the required level of accuracy.

  On the first-principles side of the wet-systems description,
ab initio molecular dynamics (AIMD) based on density-functional
theory (DFT) is the technique that, on one hand, satisfies the needs
mentioned above, and, on the other, is efficient enough to allow
for the system sizes and time scales needed in the simulations
with today's computers.
  Indeed, the last ten years have witnessed the success of many DFT-based 
AIMD simulations of liquid water, including work on structural, dynamical, 
chemical and electronic properties.\cite{Sprik96,Ortega96,Silvestrelli99jcp,
Silvestrelli99prl,parrinelloOH,parrinello-wet-e,Voth02}

  Recent reports,\cite{Grossman04,Schwegler04,Asthagiri03} however, have 
questioned some of the results of earlier studies, showing that if the
simulations are allowed to run for longer times, the diffusivity drops by 
one order of magnitude and the liquid becomes over-structured.
  The discrepancy with experiments\cite{waterexpD,waterexp1,waterexp2} is 
still unexplained. 
  Prime suspects are the fundamental limitations of present-day AIMD 
simulations of liquid water: the inability of gradient-corrected (GGA) 
density functionals to describe dispersion interactions, and the neglect 
of quantum fluctuations in the classical description of the nuclear dynamics.

  The need for long equilibration times and the poor structural and
dynamical results have been further assessed in a recent 
contribution,\cite{mvfs04} where a relaxation process of a 
time scale larger than 20~ps was found at room temperature.
  It was observed to be related to rearrangements of the 
hydrogen-bond (HB) network in the liquid, expressed by the
concentration of under-coordinated molecules, including a substantial
presence (and relevance) of bi-coordinated ones.
  Both the radial distribution functions (RDFs) and the
diffusivity were found to correlate strongly with the HB
network imperfection, in close analogy to what Giovambattista {\it et al}
\cite{Giovambattista02} found in supercooled water.
  Furthermore, while the network was slowly evolving,
the diffusivity was observed to equilibrate much faster 
to the instantaneous network state than to the final temperature.
  The mentioned paper also shows that the structure, diffusivity and network
characteristics obtained, would very nicely fit experiment if
they corresponded to an effective temperature of $\sim$240~K, instead
of the 300~K actually used in the AIMD simulations.

  All those results were obtained for gradient-corrected (GGA) 
density functionals, specifically for two proposals, 
BLYP\cite{blyp1,blyp2} and PBE.\cite{PBE}
  Both very popular and quite successful, the former comes from
the quantum-chemistry community and the latter from the condensed-matter
physics community.
  Even if they were constructed with different criteria, from a 
fundamental point of view it is good news that both functionals give 
very similar results.\cite{Grossman04}
  From a more practical view point, however, it would be highly desirable to
have a satisfactory and efficient DFT description of liquid water in order
to be able to address complex wet systems, especially since
the more technical aspects for efficient DFT calculations (including
linear scaling) have already been successfully tested for this 
system.\cite{mvfs04}

  Alternative GGAs are being explored at the moment, and very promising
results have been found by VandeVondele {\it et al.}\cite{Joost},
who tried several functionals finding much more satisfactory results.
  Since their functionals originate from the chemistry community,
where special emphasis is put on accuracy for light elements, and since
our interest in wet mineral surfaces involves heavier elements, we 
think it is worthwhile to complement their work with a similar study
with functionals proposed more generally for the whole periodic table.
  In the following we present results for the structure, diffusivity
and equilibration times for RPBE.\cite{rpbe} 
  This GGA functional is a modification of PBE, well within its philosophy and
fulfilling all its fundamental criteria.
  It is also very accessible from the points of view of coding 
(only a few lines differ from a PBE implementation) and computational
demands, remaining at the GGA level.
  It has already been tested for varied systems, including systems with heavy
elements.\cite{Eichler01,Molina03}
  Functionals beyond GGA (meta-GGAs),\cite{metagga1,metagga2} hybrid 
functionals,\cite{hybrid1,hybrid2,hybrid3} and functionals including Van der 
Waals interactions\cite{Kohn98,Lundqvist04} should also 
be checked, but their efficiency is lower, and the scope of wet systems 
that can be addressed at present is much narrower than for GGAs.
  None of these functionals will be addressed here.

  Asthagiri {\it et al.}\cite{Asthagiri03} tested an alternative 
GGA, very similar to PBE as well, called revPBE.\cite{revPBE}
  The spirit of the PBE modification of this functional is similar
to RPBE (except for the fact that the latter respects the Lieb-Oxford
bound locally, which ensures a global fulfilment of the condition
for any electronic density) representing an alternative worth considering.
  Good revPBE results for liquid water were obtained by those 
authors\cite{Asthagiri03} before the long time-scale relaxation problem was 
reported.
  Under the new circumstances it seems that the equilibration and simulation 
times used there\cite{Asthagiri03} could have been too short.
  A revision of those tests would also be useful, but is beyond the
scope of the present work.

\section{Method}

  The simulations were performed using the Kohn-Sham approach\cite{Kohn-Sham} 
to DFT\cite{Hohenberg-K} in the generalized-gradient approximation (GGA).
  The BLYP\cite{blyp1,blyp2} and RPBE\cite{rpbe} 
exchange-correlation functionals were used and are compared in the following.
  Core electrons were replaced by norm-conserving 
pseudopotentials\cite{Troullier-Martins} in their fully non-local 
representation.\cite{Kleinman-Bylander}
  Numerical atomic orbitals (NAO) of finite support were used as basis set,
and the calculation of the self-consistent Hamiltonian and overlap matrices 
was done using the linear-scaling {\sc Siesta} 
method.\cite{SiestaPRBRC,SiestaJPCM}
  Integrals beyond two-body were performed in a discretized real-space
grid, its fineness determined by an energy cutoff of 150 Ry.
  A double-$\zeta$ polarized (DZP) NAO basis set was used, which had
been obtained following the method proposed in 
Refs.~\onlinecite{javibases,edubases} for a confining 
pressure\cite{edubases} of 0.2~GPa. 
  The validation of the method, pseudopotentials and basis set can be
found in Ref.~\onlinecite{mvfs04}, including the approximations
required for linear scaling.

\begin{table}
\caption[]{AIMD simulations performed in this work, all of them
with periodic boundary conditions in a cubic cell of size $a$=9.865 \AA. 
DF stands for the particular density functional used, $T$ for the final 
equilibrated temperature, $\tau_{sim}$ for the AIMD simulation time after
AIMD equilibration, $\tau_{eq}$ for the AIMD equilibration time,
``Model" for the model used for preparation, $T_{pre}$ for the
temperature at which the preparation model had been equilibrated and
$T_i$ for the AIMD initial temperature (after the $\tau_{eq}$ anneal).
Temperatures in K and times in ps.}
\begin{tabular*}{9cm}{c@{\extracolsep{\fill}}cccclcc} 
\hline \hline
\# & DF & $T$ & $\tau_{sim}$ & $\tau_{eq}$ &
Model & $T_{pre}$ &$T_i$ \\
\hline
1  & BLYP  & 298 & 20 & 4 & BLYP  & 315 &300\\
2  & BLYP  & 315 & 32 & 6 & SPC/E & 300 &300\\
3  & RPBE  & 300 & 17 & 5 & BLYP  & 315 &300\\
4  & BLYP  & 345 & 30 & 4 & TIP5P & 325 &325\\
\hline \hline
\end{tabular*}
\label{table1}
\end{table}

  We performed AIMD simulations for 32 molecules of heavy water for 
the BLYP and RPBE density functionals (a detailed comparison between 
BLYP and PBE for this system can be found in Ref.~\onlinecite{Grossman04}).
  Details are given in Table~\ref{table1}.
  AIMD equilibration is achieved by means of temperature annealing
(velocity re-scaling),\cite{allen-tildesley} while the actual simulations are
performed by Verlet's integration.\cite{allen-tildesley}
  In all the simulations the time step used was 0.5 fs.
  The observed total-energy drifts corresponded to drifts in the system
temperature between 0.26 K/ps and 0.36 K/ps.
  The simulations were performed at constant volume, {\it i.e.}, for fixed 
cell size and shape, under periodic boundary conditions.
  As in Ref.~\onlinecite{mvfs04}, the time scale of the HB relaxation
process is monitored by following its non-equilibrium behaviour.
  The ``moving time window" used in Ref.~\onlinecite{mvfs04} is
used to obtain the time dependence of the variables of interest.

  Empirical simulations were performed using different force fields
(TIP5P\cite{tip5p} and SPC/E\cite{SPC/E}) in order to prepare reasonably
equilibrated starting points for AIMD.
  These simulations were performed with the GROMACS MD
package\cite{gromacs1,gromacs2} under constant volume and temperature
conditions using a Berendsen-type thermostat.\cite{berendsen}
  The empirical simulations were equilibrated during 200~ps.

\section{Results and discussion}

\begin{figure}[t!]
\includegraphics*[scale=1.5]{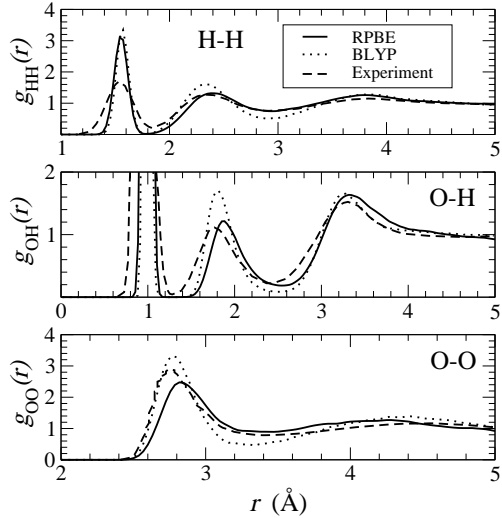}
\caption[]{Comparison of the H--H, O--H, and O--O radial distribution 
functions as obtained in this work for 32 molecules with RPBE (solid line), 
with BLYP (dotted line) and by experiment\cite{waterexp1,waterexp2}
(dashed line), at a temperature of 300~K.}
\label{grcomp}
\end{figure} 

  Fig.~\ref{grcomp} compares the RDFs obtained with RPBE and BLYP
to experiment.
  Looking mainly at the inter-molecular features, the RDFs for RPBE 
are considerably less structured than the ones for BLYP, even if still 
over-structured with respect to experiment.
  The comparison in Ref.~\onlinecite{Grossman04} between PBE and BLYP
gave much closer results.
  The too large distance for the first peak position of the O--O RDF is 
worth noting for RPBE, however.
  It indicates too weak HBs that could account for the more fluid
character of the RPBE liquid as compared with BLYP or PBE.
  The fact that this O--O distance is longer than the experimental
value (possibly weaker HB) but the liquid remains still slightly
overstructured otherwise, suggests the presence of a more fundamental 
error (possibly the same as for BLYP and PBE) that is partly compensated 
by this weaker HB.

  The values for the corresponding diffusivities are shown in
Table~\ref{table2}.
  The table also shows the effective temperatures\cite{mvfs04} for the 
different simulations, {\it i.e.}, the temperature at which the diffusivity
of the real liquid equals the AIMD value.
  The table shows the effective temperature with respect to both
normal and heavy water.
  In Ref.~\onlinecite{mvfs04} there was no need to refer to the difference
between heavy and light water, since the differences in diffusivity between 
the two ($\sim$25\%) are negligible when compared with the deviation of AIMD 
(BLYP) with respect to experiment.\cite{mvfs04}
  In this study, however, the RPBE description is much closer to
experiment, and the distinction becomes relevant.

\begin{table}
\caption[]{Diffusivity ($D$) obtained in simulations of heavy water 
for different functionals and in experiment, all at room temperature. 
Corresponding effective temperatures with respect to light water 
($T_{eff}^{\rm H_2O}$) and heavy water ($T_{eff}^{\rm D_2O}$).}
\begin{tabular*}{12cm}{l@{\extracolsep{\fill}}cll}
\hline \hline
 & $D$ ($\times 10^{-5}$ cm$^2$/s)&$T_{eff}^{\rm H_2O}$ (K)&
$T_{eff}^{\rm D_2O}$ (K)\\
\hline
BLYP             & 0.20 & 240 (--20\%) & 250 (--17\%) \\
BLYP$^a$         & 0.15 & 237 (--21\%) & 245 (--18\%) \\
PBE$^a$          & 0.16 & 238 (--21\%) & 246 (--18\%) \\
RPBE             & 1.50 & 283 ( --6\%) & 292 ( --3\%) \\
Exp (D$_2$O)$^b$  & 1.87 &  & \\
Exp (H$_2$O)$^b$  & 2.30 &  & \\
\hline \hline
\end{tabular*}
\footnotetext{
$^a$Ref.~\onlinecite{Grossman04};
$^b$Ref.~\onlinecite{waterexpD};}
\label{table2}
\end{table}

\begin{figure}[t!]
\includegraphics*[scale=1.5]{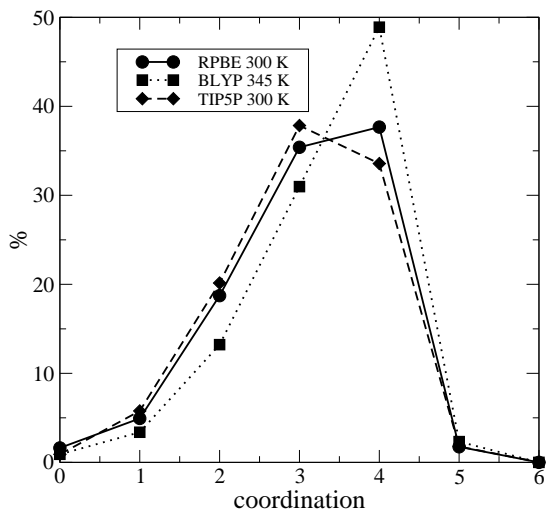}
\caption[]{Distribution of molecules with different coordinations.
Room-temperature RPBE (circles, solid line) is compared with 
the TIP5P potential\cite{tip5p} (diamonds, dashed line) at 
room temperature and with BLYP (squares, dotted line) at 345 K.}
\label{coord-hist}
\end{figure}

  As shown in Ref.~\onlinecite{mvfs04}, several results for the
room-temperature simulations fairly reproduce results of experiments
or empirical models at the effective temperature.
  Besides the diffusivity, features of the RDFs like the heights
of the first minimum and the second maximum of the O--O RDF would
conform to this temperature scaling.
  The HB network also follows this scaling.
  In Fig.~\ref{coord-hist} the distribution of molecular coordinations is
compared for RPBE and TIP5P at room temperature and BLYP at 345~K.
  The RPBE histogram shows a very slightly more over-coordinated liquid 
than the TIP5P one, suggesting a slightly lower effective temperature 
than 300~K. 
  A remarkable agreement appears when comparing the diffusivities
of both simulations, BLYP at 345~K and RPBE at 300~K.
  As reported in Ref.~\onlinecite{mvfs04} the diffusivity of the former
is $\sim 1.5\times 10^{-5}$ cm$^2$/s, identical to the one found for 
the latter in this work.

\begin{figure}[t!]
\includegraphics*[scale=1.5]{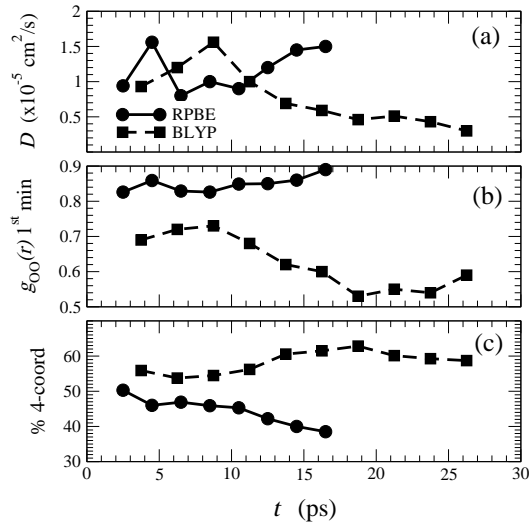} 
\caption[]{Evolution of the diffusivity (a), the height of the first minimum
of $g_{\rm OO}(r)$ (b), and the concentration of coordination defects (c), 
in simulation 3 (RPBE) calculated in 5-ps-windows every 2 ps (solid line, 
circles) and in simulation 2 (BLYP) calculated in 7.5-ps-windows every 2.5 ps 
(dashed line, squares).}
\label{evolution}
\end{figure}

  Fig.~\ref{evolution} shows the evolution of the diffusivity
and relates it to the evolution of the liquid structure, as monitored by
the heights of the first minimum and of the second maximum of $g_{\rm OO}(r)$.
  The ``moving-window" approach defined in  Ref.~\onlinecite{mvfs04} has been
used to compute the time averages. 
  In the RPBE simulation windows of 5~ps were considered every 2~ps, 
whereas in the BLYP simulation the windows were of 7.5~ps every 2.5~ps, 
given the faster evolution of the former.
  Even if none of the trajectories has strictly evolved into a stationary
situation, the figure reveals that the characteristic time of RPBE
is shorter than the one of BLYP.
  However, from the tentative assimilation of the experimentally measured
relaxation times\cite{Sette04} to our equilibration times, one would expect
for an effective temperature of 292~K a substantially shorted relaxation 
time.
  Surprisingly, it does not seem to be the case here, our equilibration 
time scale being rather on the 10~ps range or longer.

\section{Conclusions}

  We have performed DFT-based AIMD simulations of room-temperature
liquid water, comparing the characteristics of the fluid obtained
by BLYP and RPBE.
  The main conclusions are summarised as follows.

  $(i)$ The RDFs for RPBE are closer to experiment than
the corresponding ones for BLYP.
  RPBE water is less structured than BLYP, still slightly
over-structured with respect to experiment.

  $(ii)$ The diffusivity obtained for RPBE is still too low, but
much closer to experiment than the value obtained for BLYP, 
with a factor of two underestimation instead of one order of
magnitude.

  $(iii)$ For room-temperature AIMD, the effective temperature (with
respect to heavy water experimental results) is $\sim$292~K
(--3\%) for RPBE, while for BLYP it is $\sim$250~K (--17\%).
  In addition to the quantitative advantage of performing RPBE
simulations there is a qualitative advantage in that neither of these RPBE 
simulations would be for supercooled water.

  $(iv)$ The AIMD HB-network at AIMD temperature is extremely
similar to the TIP5P network at the effective temperature.

  $(v)$ The equilibration time-scale is around 10~ps or longer for 
room-temperature RPBE, as compared with more than 20~ps for BLYP.
  The fact that it is not substantially shorter is somehow 
surprising.\cite{mvfs04}

  Even if there are reasons to believe that the good results 
found for RPBE are not due to fundamental reasons, but rather to
error cancellations, still, the possibility of having a 
well-behaved GGA water is appealing.
  We consider that it is worth further exploring the performance
of RPBE in other wet systems, possibly starting from well characterised
ion hydration shells.

\acknowledgments

  We thank M. Sprik and J. VandeVondele for useful discussions.
  We acknowledge financial support from the British Engineering and
Physical Sciences Research Council, the Cambridge European Trust,
and the Comunidad Aut\'onoma de Madrid.
  The calculations were performed in the Cambridge Cranfield 
High Performance Computing Facility.

\bibliographystyle{nature}
\bibliography{mvf} 

\end{document}